\documentclass[12pt,psfig]{iopart}
\usepackage{iopams}
\usepackage{epsf}
\usepackage{euscript}
\usepackage{graphicx}
\renewcommand{\vec}[1]{{\ensuremath{\mathbf{#1}}}}

\begin{document}
\title {Geometric characterization of nodal domains: the area-to-perimeter ratio}
\author{Yehonatan Elon$^1$, Sven Gnutzmann$^{2,3}$,
  Christian Joas$^3$ and Uzy Smilansky$^1$}
\address {$^1$Department of Physics of Complex Systems,\\
 The Weizmann Institute of Science, 76100 Rehovot, Israel\\
$^2$School of Mathematical Sciences,\\
 University of Nottingham, Nottingham NG7 2RD, United Kingdom\\
$^3$Fachbereich Physik,\\
 Freie Universt\"at Berlin, 14195 Berlin, Germany}
\begin{abstract}
In an attempt to characterize the distribution of forms and shapes
of nodal domains in wave functions, we define a geometric parameter
- the ratio $\rho$ between the area of a domain and its perimeter,
measured in units of the wavelength $1/\sqrt{E}$. We show that the
distribution function $P(\rho)$ can distinguish between domains in
which the classical dynamics is regular or chaotic. For separable
surfaces, we compute the limiting distribution, and show that it is
supported by an interval, which is independent of the properties of
the surface. In systems which are chaotic, or in random-waves, the
area-to-perimeter distribution has substantially different features
which we study numerically. We compare the features of the
distribution for chaotic wave functions with the predictions of the
percolation model to find agreement, but only for nodal domains
which are big with respect to the wavelength scale. This work is
also closely related to, and provides a new point of view on
isoperimetric inequalities.
\end{abstract}

\section{Introduction}\label{sec:introduction}
In this work we study the (real) eigenfunctions of the
Laplace-Beltrami operator $-\Delta _{\mathcal{M}}$ on a Riemannian
surface $\mathcal{M}$ with Dirichlet boundary conditions (if
$\mathcal{M}$ has boundaries). Consider a real eigenfunction which
satisfies
\begin{eqnarray}
 \label{eq:helmholtz}
\left(\Delta_{\mathcal{M}}+E_j\right)\psi_j(r)=0 \quad , \quad
\left. \psi_j(r) \right|_{r\in \partial \mathcal{M}} =0 \ .
\end{eqnarray}
The nodal domains are the maximally connected domains in
$\mathcal{M}$ where $\psi_j$ has a constant sign. The nodal set (or
the set of nodal lines) is the zero set:
$U_j=\{r\in\mathcal{M}:\psi_j(r)=0\}$, which also forms the
boundaries of the nodal domains. We shall denote the nodal count
(i.e. the number of nodal domains) of $\psi_j(r)$ by $\nu_j$.\\
The investigation of quantum signatures of 
classical chaos and integrability has been a hot 
topic in
quantum chaos for a long 
time \cite{Stockmann,Haake}.
In the past few years, the interest in nodal domains, their counting
and their morphology increased 
after Blum \emph{et al} \cite{Blum}
proposed a quantitative method which 
distinguishes between the
distributions of nodal counts in 
domains where the underlying
classical dynamics is integrable (separable) or chaotic. 
This added a new approach, the statistical
investigation of nodal patterns, to the 
more common investigation methods
of spectral or
wavefunction statistics, which are
often connected to random-matrix 
theory \cite{Guhr}.
Blum \emph{et al} showed
that if $\nu_j$ is the nodal count of the $j$th energy eigenstate of
a domain , then $\xi_j=\nu_j/j$ has a limiting distribution
$P(\xi)=\lim_{n\rightarrow\infty}\frac1n\sum_{j=1}^n
\delta(\xi-\xi_j)$ , where the characteristics of the distribution
depend on the classical properties of the domain. For separable
domains, $P(\xi)$ has a square root singularity at a
(system-dependent) maximum value, while for chaotic systems $P(\xi)$
is (approximately) normally distributed. Comparison between the
numerical results for chaotic billiards and the random-wave ensemble
supports Berry's conjecture \cite{Berry77} - wave functions in a
chaotic system behave in the limit of high energy like a random
superposition of plane waves. By that, the qualitative observation
of Miller \emph{et al} \cite{Handy}, that nodal sets can be used to
distinguish between wave functions in chaotic and integrable
domains, could be tested in a quantitative way. Other studies of
various quantities - which pertain to the morphology and complexity
of the nodal network - were published in the mathematical and
physical literature, building upon the older results regarding the
bounds on the total lengths of the nodal lines and their curvature
\cite {Donnelly,Bruning}. E.g., in \cite{Berry02} the distribution
of the curvature is calculated, in addition to the mean and the variance
of the total length of the nodal set. The distribution of the
avoidance distances between nodal lines was also computed
\cite{Monastra}, to
mention few examples.\\
An important breakthrough has been achieved by Bogomolny and Schmit \cite{Bogomolny02}
who implemented a critical percolation model that
explains the large scale structure of nodal domains in chaotic wave functions. 
This model is
supported by a variety of numerical 
calculations. For example: the
expectation value and variance of the nodal count for chaotic
billiards, as well as the distribution of areas of nodal domains,
follow the predictions of the model \cite{Blum,Bogomolny02}; The
nodal lines in the high energy limit seem (on large scales) to be
SLE$_6$ curves \cite{keating,williams,Bogomolny06} as it is proved for the boundaries
of percolation clusters \cite{Smirnov}. 
Despite the good agreement,
the percolation description is a priori 
insensitive 
to the structure of the
nodal set in scales of the order 
of a wavelength. 
In addition, it
was demonstrated by Foltin \emph{et al} \cite{Foltin04} that there
are some special measures 
with a scaling behavior which is different
for percolation and the nodal set of the 
random-wave ensemble. The latter special 
measures, in general,  
probe subwavelength scales at two
points at a (large) distance.

In this work we suggest a new (quantum mechanical) method for the
classification of billiards according to their classical properties.  
We will discuss below in what sense the
signatures in nodal patterns differ from
the scenario known for the more common spectral and wavefunction analysis.\\
Our method provides yet another test to the conjectures by Berry
and Bogomolny. The parameter which we use in order to interrogate
the morphology of nodal lines is defined as follows - We consider
the $j$th eigenfunction of (\ref{eq:helmholtz}), and its nodal
domains sequence $\{\omega_j^{(m)}\}_{m=1\dots \nu_j}$. The indices
$j,m$ specify a nodal domain; For this domain we define the
area-to-perimeter ratio $\rho_j^{(m)}$ by:
\begin{eqnarray}\label{eq:rhodef}
&&\rho_j^{(m)}={\mathcal{A}_j^{(m)}\sqrt{E_j}\over{L_j^{(m)}}}\ .
\end{eqnarray}
where $\mathcal{A}_j^{(m)}$ and $L_j^{(m)}$ are the area and
perimeter of $\omega_j^{(m)}$ and the ratio is measured in units of
the wavelength $1/\sqrt{E}$. We shall define for different ensembles
two different probability measures on the parameter
$\rho$.\\
For wave functions which satisfy (\ref {eq:helmholtz}) on a compact
domain $\mathcal{M}$, we consider a spectral interval $I=[E,E+gE], \
g>0$ with $N_I = \sharp \{j:  E_j \in I\}$ , and define:
\begin{eqnarray}\label{eq:gendis1}
&&P_{\mathcal{M}}(\rho,E,g)={1\over N_I}\sum_{E_j\in I} {1\over
\nu_j}\sum_{m=1}^{\nu_j} {\delta(\rho-\rho_j^{(m)})} \ .
\end{eqnarray}
Note that in the above, the weights of nodal domains which belong to
the same eigenfunction are equal, but not necessarily the same as
the weight of domains which belong to another eigenfunction.\\
The second probability measure pertains to an ensemble of wave
functions on unbounded domains, in our case - the gaussian
random-wave ensemble (which will be described in section
\ref{sec:Random and chaotic}). Since the wave functions do not
satisfy any boundary condition, we consider them over an arbitrarily
large and fixed domain $\Omega\subset\mathcal{\mathbb{R}}^2$, and
include only the nodal domains which are strictly inside $\Omega$.
We denote their number for a given member of the ensemble by
$\nu_{\Omega}$, and define:
\begin{eqnarray}\label{eq:gendis2}
&&P_{rw}(\rho,E,\Omega)=\left<{1\over
\nu_{\Omega}}\sum_{\omega_j\subset\Omega}{\delta(\rho-\rho_j)}\right>
\end{eqnarray}
The reason for using two different measures is the different nature
of the problems at hand. However in the limit the two measures
coalesce. We shall investigate the existence and the features of a
high energy limiting distribution
\begin{eqnarray}\label{eq:limitP}
P(\rho)=\lim_{E\rightarrow\infty}P(\rho,E) .
\end{eqnarray}
The choice of the area-to-perimeter ratio $\rho$ as a parameter to
characterize the geometry of nodal domains is is inspired by the
following considerations: The nodal pattern for separable surfaces
is a checker-board, where a nodal domain $\omega$ is asymptotically
a rectangle with sides of the order of a wavelength. Therefore
$A_\omega\sim E^{-1},L_\omega\sim E^{-\frac12}$ and $\rho_\omega$
will be of the order of one. Similarly, according to the percolation
model, a nodal domain of a chaotic surface is asymptotically shaped
as a chain \cite{Bogomolny02} with $n$ cells, where for each cell
$A_{c}\sim E^{-1},L_{c}\sim E^{-\frac12}$ where $L_c$ is the cell's
contribution to the nodal domain's perimeter (see
\ref{eq:A_to_L_perco}). We get that in both cases the parameter
$\rho$ for a typical nodal domain will be of the order of unity,
yielding localized distributions for the two types of surfaces.
However, as will be shown below, the distributions differ
substantially for
systems with different classical properties.\\
In addition, the area-to-perimeter ratio $\rho$ is relevant not only
to the study of the high energy limit, but arises as a natural
parameter in the study of isoperimetric inequalities (see e.g
\cite{Freitas}). The restriction of a wave function $\psi_j$ to one
of its nodal domains $\omega_m$, is an eigenfunction of the
Laplace-Beltrami operator on the domain $\omega_m$, with Dirichlet
boundary conditions. Since it consists of a single nodal domain,
Courant theorem \cite{Courant} implies that it is the ground-state
of $\omega_m$. Therefore, knowing $\rho_j^{(m)}$ we can express the
ground-state energy in terms of the area and perimeter of
$\omega_m$. In the mathematical literature there are known bounds
for such expressions - A relevant example is the bound for convex
domains, derived by Makai \cite{Makai} (lower bound) and P\'{o}lya
\cite{Polya} (upper bound, which was generalised
to all simply or doubly
connected domains by Osserman \cite{Osserman}):
\begin{eqnarray}\label{eq:MPbound}
\frac{\pi}{4}\leq\rho_c\leq\frac{\pi}{2}
\end{eqnarray}
In order to derive a distribution function $P(\rho)$ between the
extreme values, some measure on domains should be defined. Here we
confine ourselves to well defined families of domains - those
obtained as nodal domains of a given ensemble, and due to this
restriction we are able to define the measures
(\ref{eq:gendis1}),(\ref{eq:gendis2}) and study the limiting
distribution for different classes of systems.\\
In this paper we shall examine the distribution function $P(\rho)$
for separable and chaotic domains and for the gaussian random-wave
ensemble. We will show that
\begin{itemize}
\item The limit distributions we obtain have strong ``universal``
 features. That is, they depend crucially on the type of classical
 dynamics the manifold supports, and only to a lesser extent on the
 idiosyncratic details of the actual system.
\item The limiting distribution for the random-wave ensemble is
 similar to the one for chaotic domains, as predicted by Berry's
 conjecture.
\item The limiting distribution 
  for random-waves (chaotic billiards)
  is consistent with
  the percolation model but contains
  (universal) information beyond percolation
  as short length scales
  on the order of a wavelength are probed
  for small nodal
  domains.
\end{itemize}
The limitation to (quantum mechanically)
separable systems is due to the checkerboard
structure in the nodal patterns of their 
wavefunctions. A generalization to all 
integrable or pseudo-integrable domains,
where, in general, the checkerboard 
structure is lost, would be desirable.   
This highlights a general difference 
in the scenario known e.g.~from spectral
statistics where all integrable
systems (separable or not) share the same
(Poissonian) statistics. Quite contrary
statistical properties of nodal domains
in integrable systems are very different for
seperable systems with a checkerboard structure
and non-separable systems with generically
no nodal crossings \cite{Monastra,Aiba}.
Note, that when a separable system is 
slightly perturbed, all nodal
crossings will open (in an often highly
correlated way which makes the introduction
of a percolation model at this point quite
difficult) and statistical properties
will change singularly (again in contrast 
to what is known from spectral statistics
where a small perturbation smoothly 
changes the statistics).

\section{The limiting distribution of the area-to-perimeter ratio for separable domains}\label{sec:Separable} 

As was
mentioned above, the nodal network of eigenfunctions of separable
surfaces has a checkerboard structure. This follows from the fact
that one can always choose a basis in which all the eigenfunctions
can be brought into a product form. Therefore, one might expect that
the main features of the  limiting distributions of different
surfaces will be similar. We will show that this is the case, and
therefore we begin this section by explicit calculation of
$P_{rec}(\rho)$ for a rectangular billiard. The discussion of this
simple example will pave the way to computing $P(\rho)$ for other
systems (the disc billiard and a family of surfaces of revolution)
and to the identification of some common features which we assume to
be universal for all the separable systems. The
detailed computations are presented in \ref{app:a}.\\

\noindent The Dirichlet eigenfunctions for a rectangular billiard
with side lengthes $a,b$ are:
\begin{eqnarray}\label{eq:psirec}
\psi_{mn}(x,y)=\sin\frac{\pi mx}{a}\sin\frac{\pi ny}{b}\equiv
\psi_m(x)\psi_n(y)
\end{eqnarray}
The corresponding eigenvalues are:
\begin{eqnarray}
E_{mn}=\pi^2\left[\left(\frac ma\right)^2+\left(\frac nb\right)^2\right] \equiv
E_m+E_n
\end{eqnarray}
where $E_m,E_n$ can be interpreted classically as the energy stored
in each degree of freedom (A formal definition can be found in
\ref{app:a}). The nodal domains are rectangles of size:
${{\pi}\over{\sqrt{E_m}}}\times{{\pi}\over{\sqrt{E_n}}}$, therefore:
\begin{eqnarray}\label{eq:rhorec}\nonumber
\mathcal{A}_{mn}={{\pi^2}\over{\sqrt{E_m E_n}}}\quad,\quad
L_{mn}=2\pi({1\over\sqrt{E_m}}+{1\over \sqrt{E_n}})
\end{eqnarray} and
\begin{eqnarray}
\rho_{mn}=\frac{\pi}{2}
\left({\sqrt{\frac{E_m}{E_{mn}}}+\sqrt{1-\frac{E_m}{E_{mn}}}}\right)^{-1}\equiv\rho\left(\frac{E_m}{E_{mn}}\right)
\end{eqnarray}
where: $\rho\left(\frac{E_m}{E_{mn}}\right)=
\rho\left(\frac{E_n}{E_{mn}}\right)$. The mere form  of
(\ref{eq:rhorec}) implies that $\rho$ is bounded by:
\begin{eqnarray}\label{eq:rhobounds}
\frac{\pi}{\sqrt8}\leqq\rho_{mn}^{(rec)}\leqq\frac{\pi}{2}
\end{eqnarray}
Thus, the support of the distribution function $P(\rho)$ is an
interval which is narrower than (\ref{eq:MPbound}).\\
In the limit $E\rightarrow\infty$, the distribution
(\ref{eq:gendis1}) can be approximated (neglecting corrections of
order $E^{-\frac{1}{2}}$) by an integral. Performing the integration
over the variables:
\begin{eqnarray}\nonumber
k=\sqrt{E_{mn}} \quad,\quad \theta=\arctan(\frac{E_n}{E_m})
\end{eqnarray}
the distribution function is:
\begin{eqnarray} \label{eq:Prec}
\hspace{-20mm} P_{rec}(\rho)={2\over \pi}\int_0^{\pi\over
2}\delta\left(\rho-{\pi
\over{2(\sin\theta+\cos\theta)}}\right)d\theta= \left\{
\begin{array}{rcl} {4\over{\rho\sqrt{8\rho^2-\pi^2}}} &
\mbox{for} & {\pi\over{\sqrt8}}\leqq\rho\leqq{\pi\over 2}\\
0 & \ & \mbox{otherwise} \end{array}\right.
\end{eqnarray}

\begin{figure}[h]
\centering
  \includegraphics[width=4in]{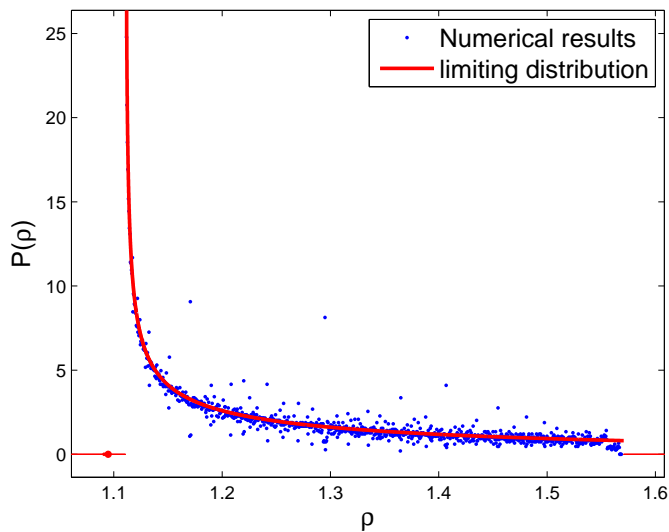}
  \caption{\label{full} The limiting distribution for a rectangular billiard (\ref{eq:Prec}) compared
  to the calculated distribution for eigenfunction with $E\cdot \mathcal{A}<10^5$}\label{fig:prec}
\end{figure}

\noindent The explicit form of $P_{rec}(\rho)$ suggests the
following qualitative and quantitative conclusions:
\begin{enumerate}
\item The existence of a limiting distribution function (which is
independent of the aspect ratio $a/b$ of the billiard) is
demonstrated.
\item It is supported by the compact interval
$\left[\pi/\sqrt{8},\pi/2 \right]$ .
\item $P_{rec}(\rho)$ is an analytic and
monotonic decreasing function in the interval where it is supported.
\item $P_{rec}(\pi/\sqrt{8}+\delta)_{_{\delta\rightarrow 0^+}}\sim\
1/\sqrt{\delta}$ .
\item $P_{rec}(\pi/2)=8/\pi^2$, Hence, $P_{rec}(\rho)$  is discontinuous
at both boundaries of the support.
\end{enumerate}
The fact that $\rho_{mn}$ depends solely on the partition of the
energy between the modes was a key element in the construction
above, and plays a similar role in computing $P(\rho)$ for the other
separable systems. When $\rho=\pi/\sqrt{8}$, we get from equation
(\ref{eq:rhorec}) that $E_m=E_n$, while $\rho=\pi/2$ means that the
energy is concentrated completely in one degree of freedom. The
concentration of probability near $\rho=\pi/\sqrt{8}$ shows that
equal partition of energy is prevalent among the nodal domains.

An explicit derivation of the limiting distribution $P(\rho)$ for
the family of simple surfaces of revolution (following Bleher
\cite{Bleher}) can be found in \ref{app:a}, in addition to a
separate derivation for the disc billiard. In both cases it is
proven that:
\begin{eqnarray}\label{eq:pofrdsor}
P_{sep}(\rho)=P_{rec}(\rho)\cdot T(\rho)
\end{eqnarray}
Where $P_{rec}$ is given by (\ref{eq:Prec}), and $T(\rho)$ is a
finite, positive and smooth function of $\rho$. Therefore the
features which characterize $P_{rec}(\rho)$ dominate $P_{sep}(\rho)$
for all the systems considered; Thus $P_{sep}(\rho)$ is supported on
the same interval and demonstrates the same type of discontinuities
at its boundaries.\\
Following the striking similarity of the distributions for all of
the investigated manifolds, we suggest that properties (i)-(v) of
$P_{rec}(\rho)$ which were derived for the rectangular billiard, are
universal features of $P_{sep}(\rho)$ for all two-dimensional
separable surfaces. We support this assumption by a heuristic model
which is presented in \ref{app:a}.

\begin{figure}[h]
\centering
\includegraphics[width=4in]{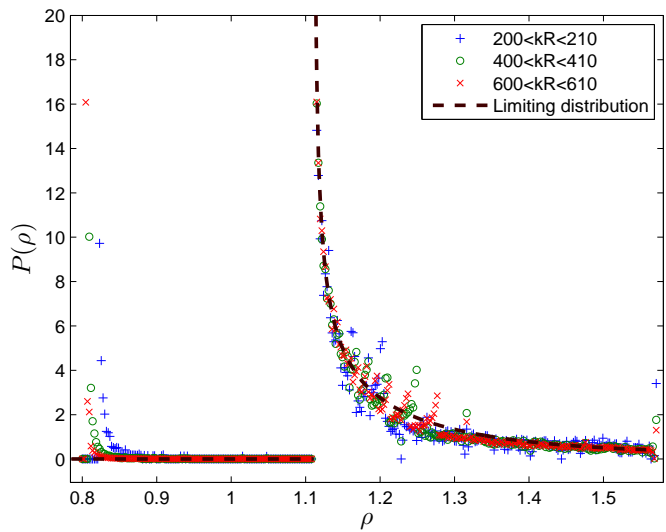}
\caption{Semi-Classical approximation of the limiting distribution
for the  disc billiard compared with numerical results for 3
different energy intervals: $200<kR<210$, $400<kR<410$ and
$600<kR<610$ where $k\equiv\sqrt{E}$}\label{fig:pdisc}
\end{figure}

\noindent Numerical simulations for the rectangle and the disc
billiards for several energy intervals show good agreement with the
analytic derivation (see figures \ref{fig:prec},\ref{fig:pdisc}).
For numerically obtained results at finite energies, two kinds of
deviations from the limiting distributions can be observed:
\begin{enumerate}
\item Fluctuations along the entire range of $\rho$ (for the disc) or discrete
jumps (for the rectangle) in the value of $P(\rho)$, which vanish in
the limiting distribution due to the convergence of the corrections
to the semi-classical approximations (i.e. turning sums over quantum
numbers into integrals and the neglect of terms of order
$E^{-\frac12}$).\item Cusps near $\rho=\frac\pi 4$ and
$\rho=\frac\pi 2$ for the disc: the origin for the appearance of
these features is due to nodal domains with exceptional geometry -
the inner domains of all wavefunctions (which are asymptotically
triangles) for the former, the domains of $\Psi_{n,0}$ (which are
ring shaped) for the latter.
\end{enumerate}
It is verified (analytically and numerically) that these differences
converge to zero as $E^{-\frac 12}$, or faster.\\

\section{The limiting distribution of the area-to-perimeter ratio for the random-wave ensemble and chaotic domains}\label{sec:Random and chaotic}

While for chaotic wavefunctions there is no known analytic
expression for the nodal lines, we will use known results about the
morphology of the nodal set in order to propose some physical
arguments for the expected distribution. The explanations we propose
are all in agreement with numerical simulations - a detailed
information about the numerical techniques and the reliability of
the results can be found in appendix B.\\
A frequently used model for eigenfunctions in a chaotic billiard is
that of the Gaussian random-wave ensemble. This is based on a
conjecture by Berry \cite{Berry77} that eigenfunctions of a chaotic
billiard in the limit of high energies have the same statistical
properties as the Gaussian random-wave ensemble.\\
A solution $\psi$ for the Helmholtz equation (\ref{eq:helmholtz})
with a given energy $E=k^2$ on a given domain, can be written as a
superposition of functions $\{\psi_l(\vec{r})\}_{l=-\infty}^\infty$
which span a complete basis, for example:
\begin{equation}\label{eq:randomwave}
\psi(\vec{r})=\sum_{l=-\infty}^\infty a_l J_l(kr)e^{il\phi}
\end{equation}
Since the solutions of (\ref{eq:helmholtz}) are real, we are
restricted (for this choice of basis) by $a_{-l}=(-1)^l a_l^*$.
According to Berry's conjecture, expanding the eigenfunctions of
chaotic billiards (in the high energy limit) in terms of
(\ref{eq:randomwave}), the coefficients $a_l$ distribute for $l\geq
0$ as independent gaussian random variables with
\begin{equation}
  \langle a_l a_{l'}^* \rangle= \delta_{l,l'}
\end{equation}
and therefore can be modeled statistically by this ensemble of
independently distributed Gaussian random-waves.\\

\begin{figure}[h]
\centering
  \includegraphics[width=4in]{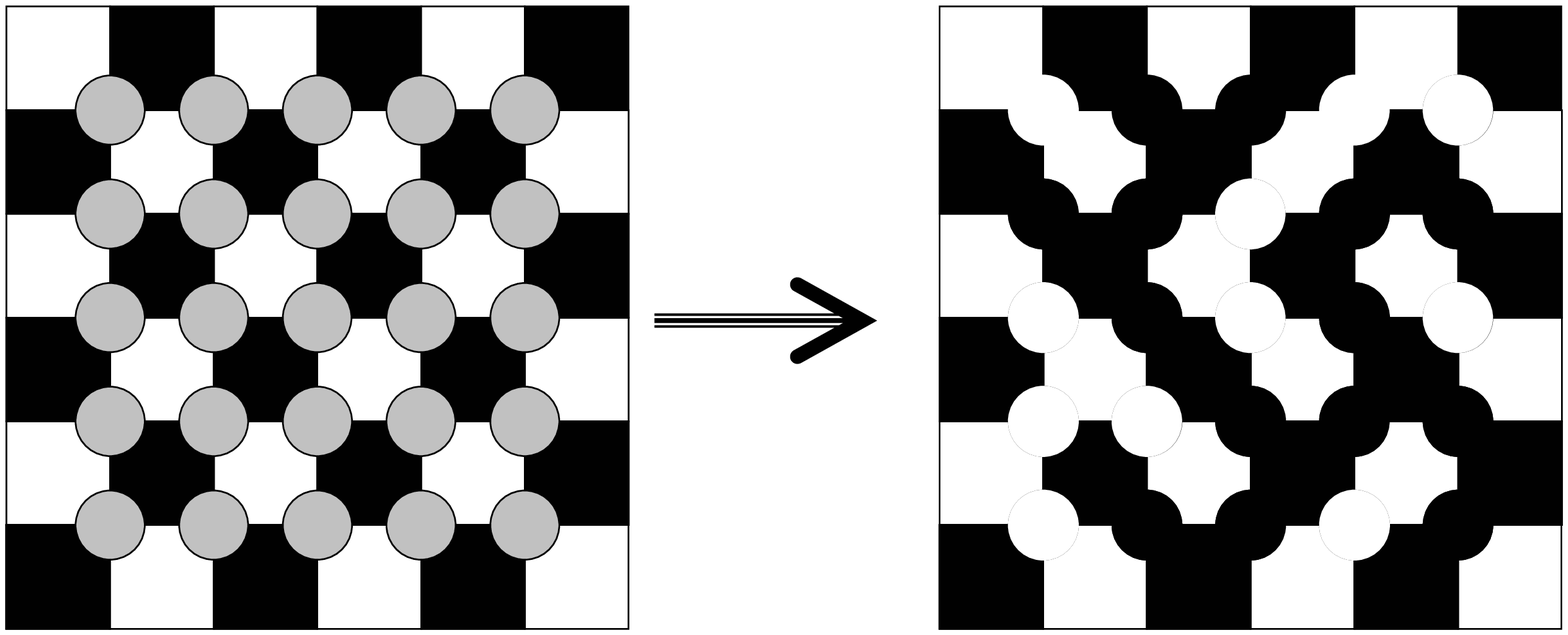}
\nonumber\includegraphics[width=3in]{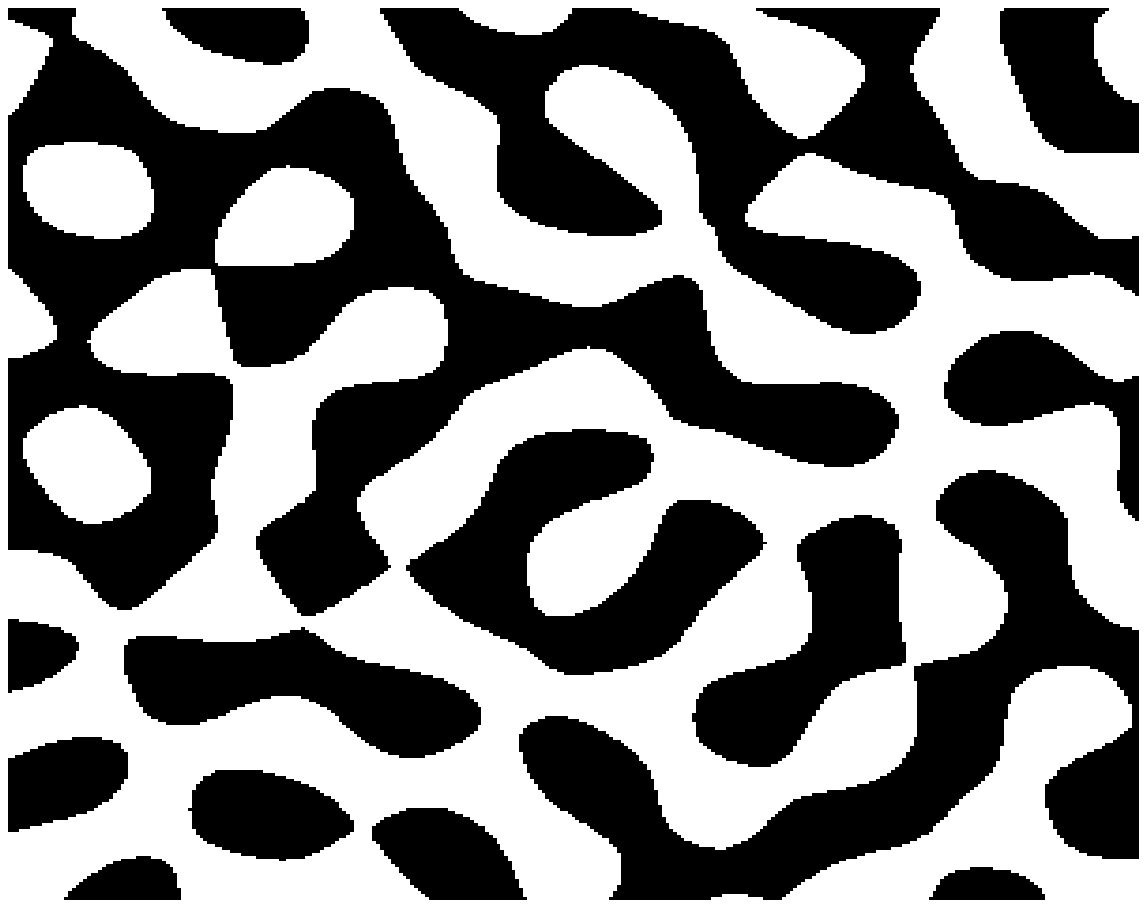}
\caption{Above - a realization of the bond-percolation model by Bogomolny and Schmit. The
dark sites are positive, and the bright are negative; at each
junction there is a saddle which can be positive or negative with
equal probability, connecting by that two of its neighbors to the
same cluster. Below - a typical realization of the nodal structure
of a random-wave on the wavelength
scale.}\label{fig:bond_percolation}
\end{figure}

\begin{figure}[h]
\centering
  \includegraphics[width=3.5in]{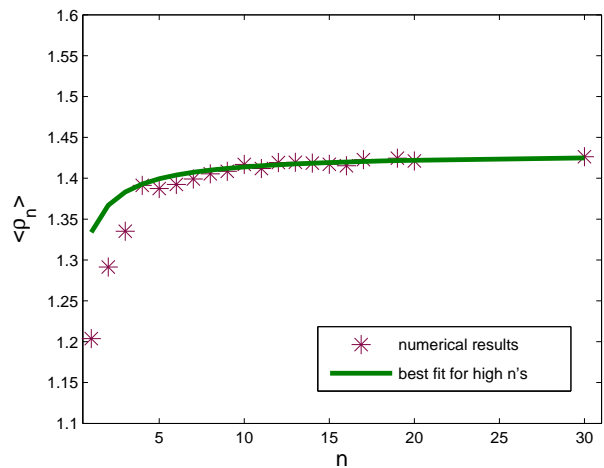}
\caption{The measured value of $\langle\rho_n\rangle$ compared to
the best numerical fitting (for high values of $n$) to (\ref
{eq:A_to_L_perco}): $\langle\rho_n\rangle=\sqrt{2}(n+0.805)/
(n+0.936)$.}\label{fig:rho_of_n}
\end{figure}

As suggested by Bogomolny and Schmit 
\cite{Bogomolny02}, the nodal domains of a random
wave are shaped as critical percolation 
clusters (see Figure 
\ref{fig:bond_percolation}), where each site 
is of an
average area
\begin{eqnarray}\label{eq:A_s}
\mathcal{A}_s=\frac{2\pi^2}{k^2}
\end{eqnarray}
where, as before, $k\equiv \sqrt{E}$. The area (or alternatively,
the number of sites) of the nodal domains (see e.g. \cite{Stauffer})
distributes (asymptotically) as a power law:
\begin{eqnarray}\label{eq:Fisherexp}
p(n)\propto n^{-\tau}
\end{eqnarray}
Where (for 2d percolation): $\tau=187/91$. For bond-percolation
model over a lattice (as illustrated in fig.
\ref{fig:bond_percolation}), the area of a cluster which spreads on
$n$ sites is $(a_1+a_2)n-a_2$\quad where $a_1$ is the area of a
single site and $a_2$ is the area of the connection between two
sites; the average perimeter is $(l_1+l_2)n-l_2$\quad where
$l_1,l_2$ are the average contributions to the perimeter of a site
and a connection (Since a cluster may contain loops which affect its
perimeter, we must speak about average). The average
area-to-perimeter ratio can be written as
\begin{eqnarray}\label{eq:A_to_L_perco}
\frac
AL=\mathcal{C}_1\left(1-\frac{\mathcal{C}_2}{n+\mathcal{C}_3}\right)
\end{eqnarray}
where $\mathcal{C}_1,\mathcal{C}_2,\mathcal{C}_3>0$. This relation
can be used as a guideline to the desired distribution
of $\rho$.\\
Indeed, as was confirmed numerically, the distribution $P(\rho)$ for
nodal domains follows (\ref{eq:A_to_L_perco}) in several aspects. We
have examined the restricted distribution for nodal domains with a
given number of sites - we define $P^{(n)}(\rho)$ to be the
distribution for nodal domains with area
\begin{eqnarray}\label{eq:A^n}
(n-\frac12)\mathcal{A}_s<\mathcal{A}^{(n)}\leq
(n+\frac12)\mathcal{A}_s
\end{eqnarray}
We found out that $P^{(n)}(\rho)$ is roughly symmetric about a mean
value: $\langle\rho_n\rangle$. As in (\ref{eq:A_to_L_perco}),
$\langle\rho_n\rangle$ is increasing with $n$, and converging to a
limiting value: $\rho_{\infty}$. Since the percolation model is
assumed to provide an exact description of the system in the
high-energy limit, we expect (\ref{eq:A_to_L_perco}) to serve as
a good approximation for large domains (see fig. \ref{fig:rho_of_n}).\\
The value of $\rho_\infty$ is a direct result of a theorem by Cauchy
for the average chord length of a domain:
\begin{eqnarray}\label{eq:chord}
\langle\sigma\rangle=\frac{\pi\cdot A}{L}
\end{eqnarray}
Where $\sigma$ is the chord length. The original theorem (which was
stated for convex domains) is extended in \cite{Mazzolo}, to
include nonconvex and multiply connected domains.\\
For nodal domains of infinite size, the statistics of the average
chord length should follow that of the entire nodal set. That in
turn is known to be \cite{Berry77,Rice}:
$\langle\sigma_{RW}\rangle=\sqrt{2}\pi/k$, therefore:
\begin{eqnarray}\label{eq:rho_infinity}
\rho_\infty=\frac{\langle\sigma_{RW}\rangle\cdot k}{\pi}=\sqrt{2}
\end{eqnarray}
The value of $\langle\rho_1\rangle$ can also be estimated: as shown
in \cite{Monastra}, the single cell nodal domains are mild
deformations of a circle of radius $r=j_0/k$ (where $j_0\approx
2.405$ is the first zero of $J_0$). Therefore:
\begin{eqnarray}\label{eq:rho_1}
\langle\rho_1\rangle=\frac{\pi r^2 k}{2\pi r}=\frac{j_0}{2}
\end{eqnarray}

\begin{figure}[h]
\centering
\includegraphics[width=4.5in]{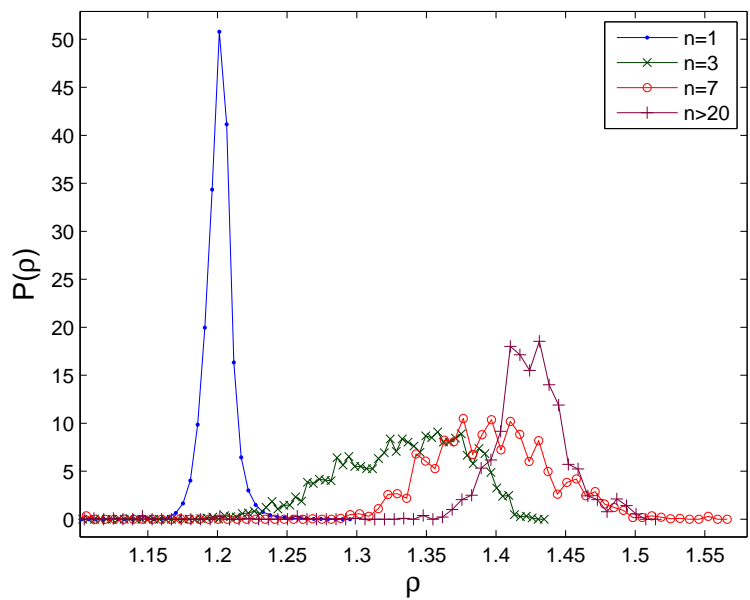}
\caption{The distribution function $P^{(n)}(\rho)$ is plotted for
several values of $n$. The mean value is increasing with $n$, while
the variance decreases (except for $n=1$) as suggested by the
heuristic model.}\label{fig:rho_per_area}
\end{figure}

\noindent In order to study the impact of the deformations on the
value of $\langle\rho_1\rangle$, we have calculated $\rho$ for a
variety of domains, like ellipses, rounded shapes with corners (a
quarter of a circle or a stadium etc.) and others. The results show
that stretching of the nodal domain (e.g.\ increasing the
eccentricity of an ellipse) increases $\rho$, while turning it
``polygonal`` (i.e.\ having points on the nodal line with very high
curvature) reduces $\rho$.\\
Derivation of ~$\langle\rho_n\rangle$~ for other values of $n$ seems
to be more complicated. However, fitting between the numerical
results for high $n$ values and (\ref{eq:A_to_L_perco}) equips us
with the empirical result (which is valid for $n\gg1$):
\begin{eqnarray}\label{eq:rho_n}
\langle\rho_n\rangle\approx\sqrt{2}\cdot\frac{n+0.805}{n+0.936}
\end{eqnarray}
Another interesting feature is the width of the distribution around
$\langle\rho_n\rangle$. Equation (\ref{eq:chord}) implies that the
variance is proportional to the variance in the average chord length
between different nodal domains of the (approximately) same area.
Therefore, the variance is expected to be smaller for larger
domains, which follows the statistics of the entire nodal network to
a larger extent. The only exception is the variance for single site
domains, which as was mentioned \cite{Monastra}, have strong
limitations on their shape, and therefore a relatively small
variation in the average chord length.\\
The bounds (\ref{eq:MPbound}) on $\rho$ should not hold in general
for the nodal domains of (\ref{eq:randomwave}); however, the
numerical bounds seem to agree with (\ref{eq:MPbound}) for all of
the measured nodal domains, including multiply connected domains,
suggesting that (\ref{eq:MPbound}) is valid for the nodal domains of
the ensemble with probability 1.

\begin{figure}[h]
\centering
  \includegraphics[width=4in]{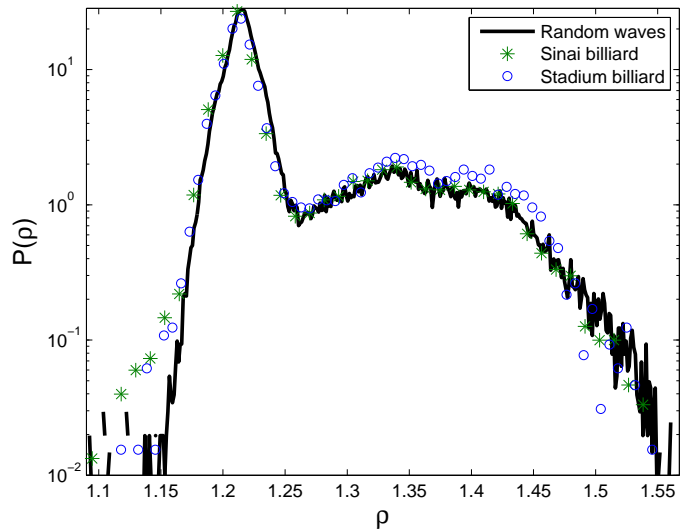}
\caption{A comparison between the distribution function $P(\rho)$
calculated for the random-wave ensemble and for inner domains of a
Sinai and stadium billiards.}\label{fig:rw_vs_inner}
\end{figure}

\begin{figure}[h]
\centering
 \scalebox{0.5}{\includegraphics[87,262][507,578]{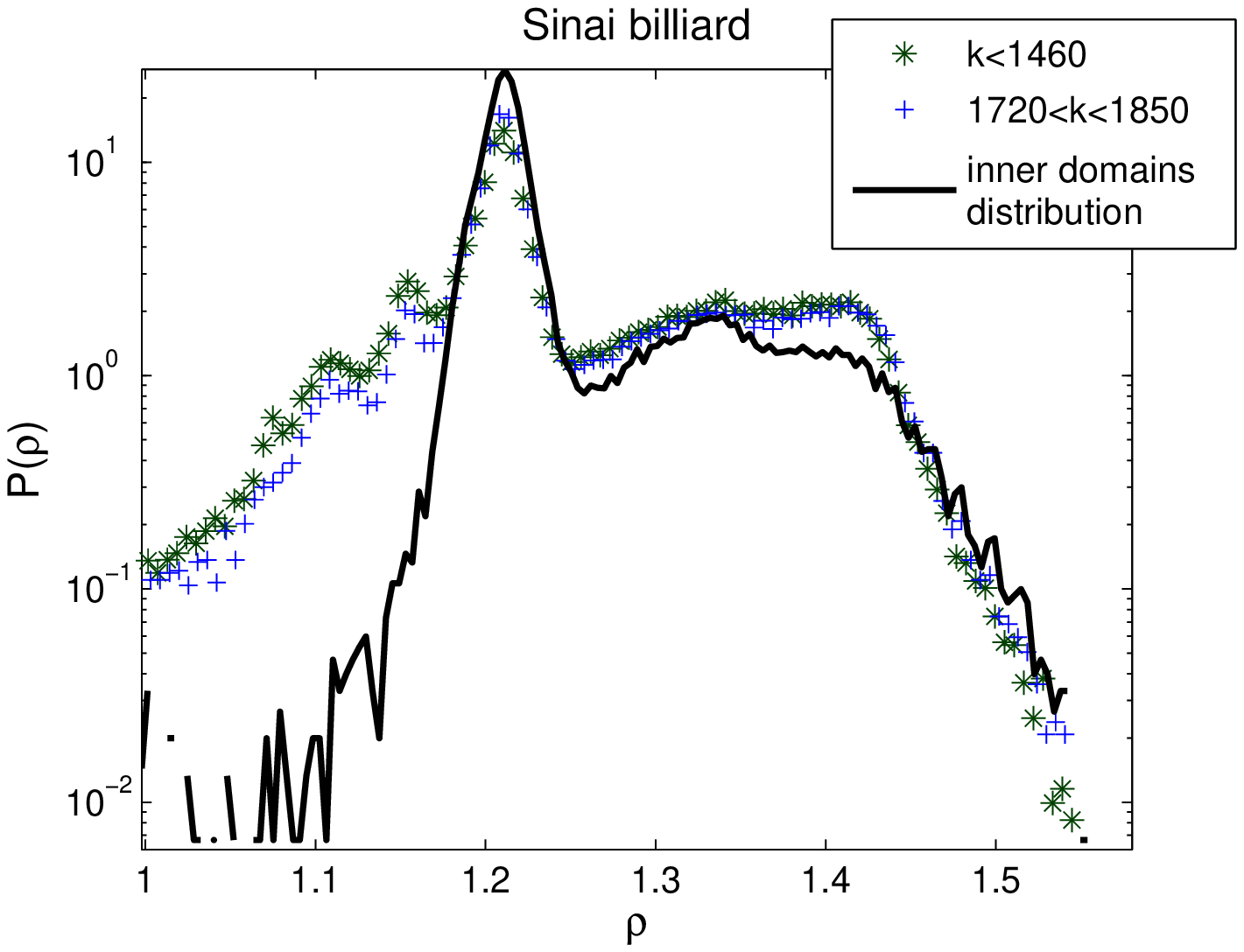}}
  \nonumber\scalebox{0.5}{\includegraphics[87,262][507,578]{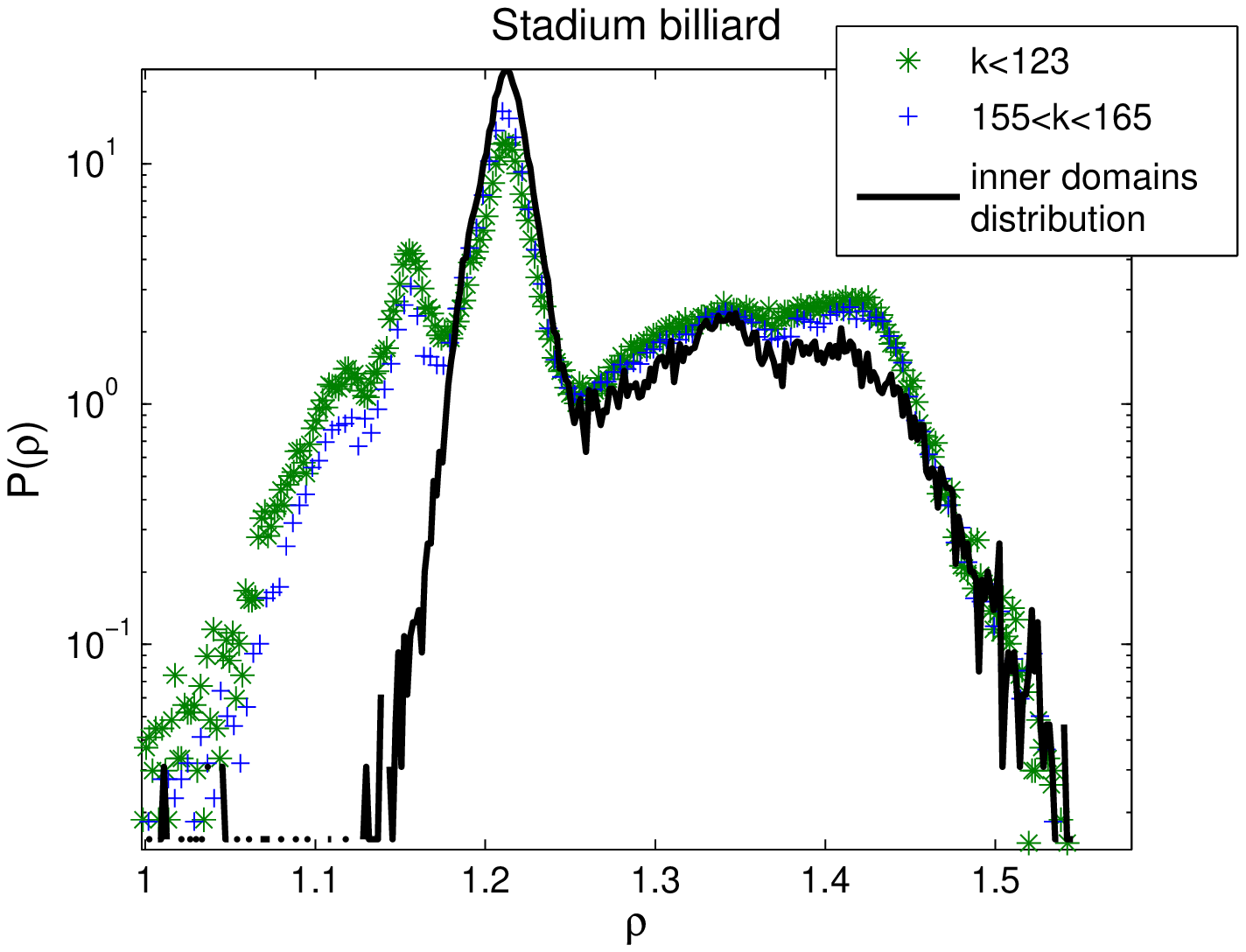}}
\caption{The distribution function $P(\rho)$ for a Sinai and stadium
billiards. We compare the distribution for relatively low energy
intervals (asterisks), higher intervals (crosses) and inner domains
only (solid line). The distribution function indeed (slowly)
converges to the distribution of inner domains (the peaks at
$\pi/\sqrt8$ and $\sqrt{2}$ are lowered, the one at $j_0/2$ is
elevated), which is similar to the distribution for the random-wave
ensemble (see fig. 11)}\label{fig:inner_vs_outer}
\end{figure}

\noindent The distribution of $\rho$ for all of the nodal domains is
given by:
\begin{eqnarray}\label{eq:ptotal}
P(\rho)=\sum_{n=1}^\infty p(n)P^{(n)}(\rho)
\end{eqnarray}
where $p(n)$ is given asymptotically by (\ref{eq:Fisherexp}).\\
Fig. \ref{fig:rw_vs_inner} shows the calculated distribution for 3
different systems - A random-wave ensemble, the inner domains of a
Sinai billiard and those of a stadium billiard. Comparing the
functions, we find additional strengthening to Berry's conjecture.
There are boundary effects of chaotic billiards - e.g. a peak in the
distribution $P(\rho)$ near $\pi/\sqrt8$ and $\sqrt{2}$, and a lower
probability for $\rho\sim j_0/2$, however they vanish in the
semi-classical limit (see fig. \ref{fig:inner_vs_outer}). In our
study it was easy to put those effects aside - if we consider only
inner nodal domains for the chaotic billiards (as in fig.
\ref{fig:rw_vs_inner}), we observe no prominent differences between
the distributions.

\begin{figure}[h]
\centering
\includegraphics[width=4in]{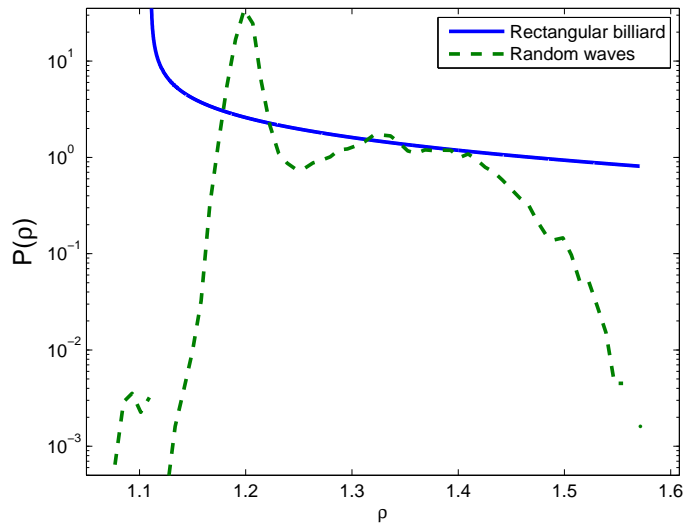}
\caption{A comparison between the limiting distribution
(\ref{eq:Prec}) derived for the rectangle billiard and the
random-wave ensemble.}\label{fig:rw_vs_separable}
\end{figure}

\section{Conclusions}
The main results of this work can be summarized as follows:
\begin{enumerate}
\item  The distribution function $P(\rho)$ of the area-to-perimeter
 ratio $\rho$, distinguishes between billiards with separable or chaotic
 classical limit (see fig. \ref{fig:rw_vs_separable}).
\item The distribution (\ref{eq:pofrdsor}) for the examined separable
 billiards has some universal features, such as a common support and a
 square-root divergence at the lower support. In all studied cases
 the distribution is a mild deformation of the distribution that we
 found for a rectangle.
\item In accordance with the random-wave conjecture, we find numerically
 that chaotic billiards (stadium and Sinai) have a universal limiting
 distribution $P(\rho)$, and it converges to the distribution found for
 the  random-wave ensemble. By considering only the inner nodal domains for
 billiards, the agreement can be shown also for finite energies.
\item The numerical results suggest that for nodal domains of a random
 wave or of eigenfunctions of chaotic billiards, the area-to-perimeter ratio
 is bounded by (\ref{eq:MPbound}), i.e. $\pi/4\leq\rho_{rw}\leq\pi/2$,
 including nonconvex and multiply connected nodal domains (for which these
 bounds have not been proven), with probability one.
\item We examined the percolation model for the nodal set of random
 waves from the perspective of the area-to-perimeter ratio. It is
 shown that on the wavelength scale the geometry of the nodal domains
 can only be poorly characterized by percolation arguments. However,
 for large domains the geometry can be described by heuristic
 expressions like (\ref{eq:A_to_L_perco}), which are consistent with
 percolation theory.
\end{enumerate}

\ack{The work was supported by the Minerva Center for non-linear
Physics and the Einstein (Minerva) Center at the Weizmann Institute,
and by grants from the GIF (grant I-808-228.14/2003), and EPSRC
(grant GR/T06872/01.) and a Minerva grant.}

\appendix
\section{Derivation of the area-to-perimeter distribution for some separable surfaces} \label{app:a}
In this appendix we suggest a heuristic model for the universal
features of the limiting distributions of the area-to-perimeter
ratio for two-dimensional separable domains. The model is supported
by an explicit derivation of the limiting distribution for the disc
billiard and for simple surfaces of revolution.

\subsection{Universal features of the distribution}
We begin by considering the classical geodesic flow in a
two-dimensional compact domain (e.g. a billiard). For a separable
domain, a trajectory can be specified by its action-variables:
\begin{eqnarray}\label{eq:action}
m=\oint p_1dq_1 \quad,\quad n=\oint p_2dq_2
\end{eqnarray}
where $q_1 ,q_2$ are the (separable) coordinates, $p_1 ,p_2$ are the
conjugated momenta and the integration is over one period of the
specified coordinate. At every point along the trajectory, the
energy $E_{mn}=\frac12\left|\vec{\dot{q}}\right|^2$ can be expressed
as $E_{mn}=E_{m}+E_{n}$, where:
\begin{eqnarray}\label{eq:partE}
E_m\equiv\frac12\left|\vec{\dot{q}}\cdot\hat{\vec{q}}_1\right|^2\quad,\quad
E_n\equiv\frac12\left|\vec{\dot{q}}\cdot\hat{\vec{q}}_2\right|^2
\end{eqnarray}
where $\hat{\vec{q}}_1,\hat{\vec{q}}_2$ are the local unit vectors -
if we consider circular domain for example, then
$\hat{\vec{q}_1}\equiv\hat{\vec{r}}=(\cos\theta,\sin\theta),
\hat{\vec{q}_2}\equiv\hat{\vec{\theta}}=(-\sin\theta,\cos\theta)$.
In general $E_m,E_n$ are not constants of motion. The only
exceptions are the trajectories in a rectangular billiard.

From a quantum point of view, the eigenstates of Schr\"odinger
equation (\ref{eq:helmholtz}) for a separable domain, can be written
as $\psi_{mn}=\psi_m(q_1)\psi_n(q_2)$, while the Laplace-Beltrami
operator can be written as: $\Delta=\Delta_m+\Delta_n$, where
$\Delta_m\psi_n=\Delta_n\psi_m=0$. This allow us to define the
quantum analogue to (\ref{eq:partE}):
\begin{eqnarray}\label{eq:partEq}
E_m(q_1,q_2)=-\frac{\Delta_m\psi(q_1,q_2)}{\psi(q_1,q_2)}\quad,\quad
E_n(q_1,q_2)=-\frac{\Delta_n\psi(q_1,q_2)}{\psi(q_1,q_2)}
\end{eqnarray}
In the semi-classical limit, the spectrum of a separable domain is
given by $\{E_{mn}|{m,n\in\mathbb{N}}\}$, where $E_{mn}$ is the
energy of the classical trajectory specified by the action-variables
$m\hbar,n\hbar$ (as emerging from Bohr-Sommerfeld quantization). In
addition the semi-classical value of (\ref{eq:partEq}) converges to
the classical value (\ref{eq:partE}) for every point in the
domain.\\

\begin{figure}[h]
\centering
  \includegraphics[width=3in]{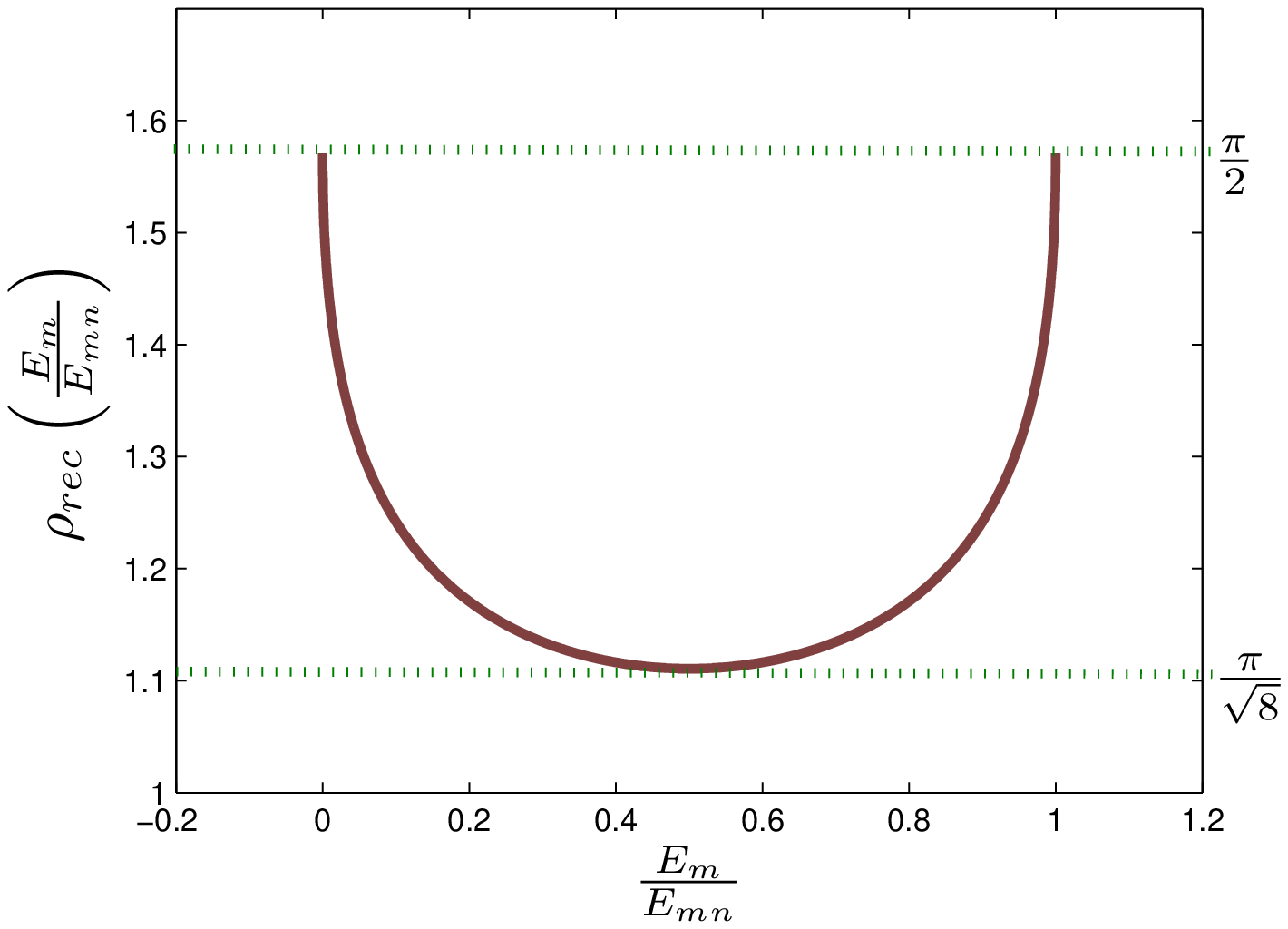}
\caption{The value of $\rho$ as a function of the partition of
 energy between the two degrees of freedom, for the rectangular billiard.}\label{fig:rho_of_ei}
\end{figure}

\begin{figure}[h]
\centering
  \includegraphics[width=3in]{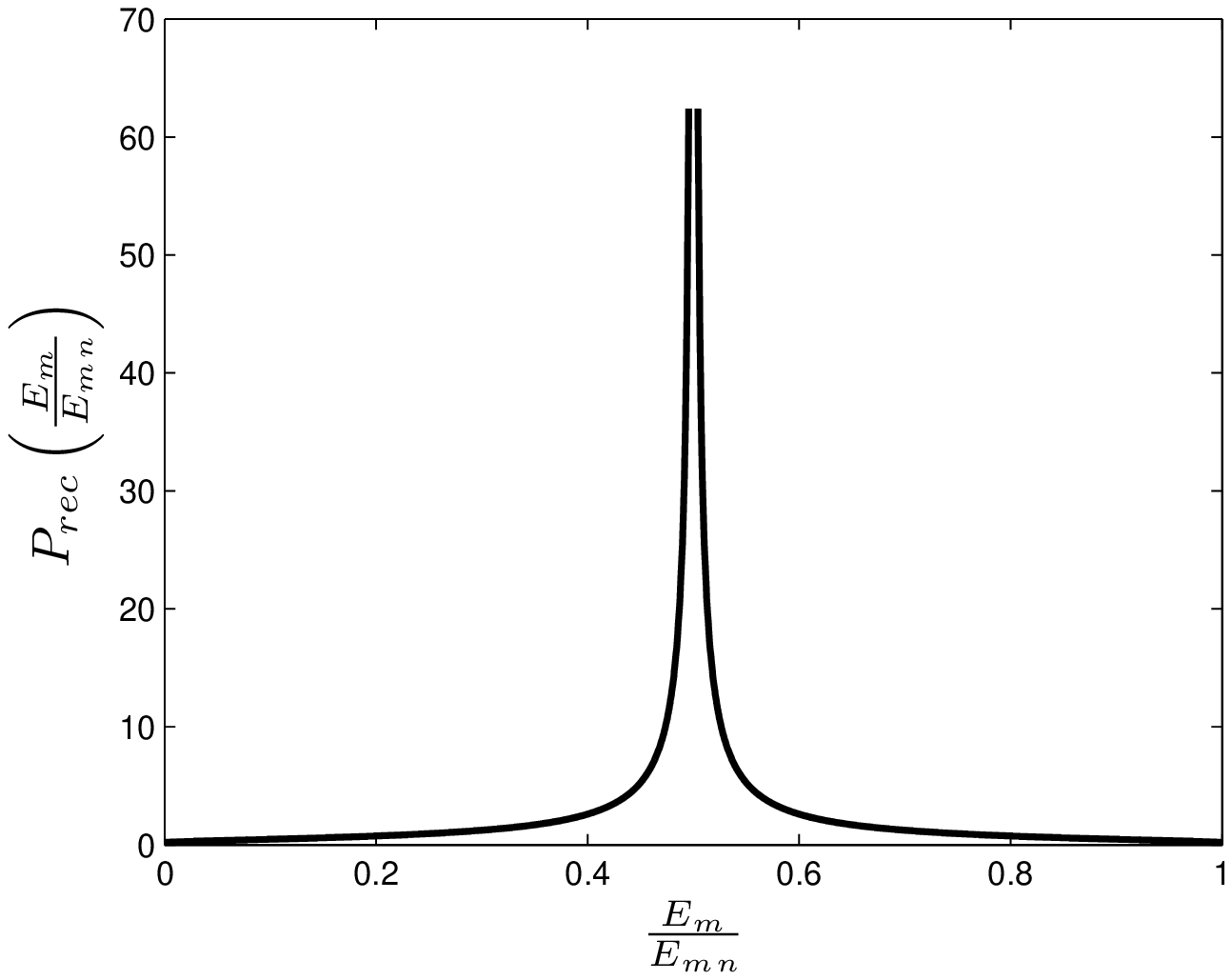}
\caption{The limiting distribution of the partition of energy
  between the two degrees of freedom for a rectangular billiard.}\label{fig:p_of_em_rec}
\end{figure}

\noindent In section \ref{sec:Separable}, the value of the
area-to-perimeter ratio for a given realization $\psi_{mn}$ for the
rectangular billiard, was derived to be
\begin{eqnarray}\label{eq:rhorec2}
\rho_{mn}=\frac{\pi}{2}
\left({\sqrt{\frac{E_m}{E_{mn}}}+\sqrt{1-\frac{E_m}{E_{mn}}}}\right)^{-1}
\end{eqnarray}
Therefore, for a rectangular billiard, the value of the (quantum)
parameter $\rho$, has also an immediate classical interpretation
(see figures \ref{fig:rho_of_ei},\ref{fig:p_of_em_rec}).\\
In order to generalize the limiting distribution which was derived
for a rectangular domain to other separable domains, we suggest the
following heuristic model:
\begin{itemize}
\item In the high energy limit, (almost all of) the nodal domains of a separable domain are
 converging to rectangles, and the wave function in the close
 neighborhood of a nodal domain is converging to (\ref{eq:psirec}).
 Therefore (in the limit) equation (\ref{eq:rhorec2}) should hold.
 However, since for general separable domains $\psi_{mn}(\vec{r})$ is not an
 eigenfunction of the operators $\Delta_m,\Delta_n$, the value of
 $\rho_{mn}^{(j)}$ for a given realization $\psi_{mn}$ will depend on the interrogated nodal
 domain $\omega_j$ .
\item In the continuum limit, the limiting distribution is of the form:
 \begin{eqnarray}\label{eq:continuum}
 P(\rho)=\frac{1}{N_I}\int_I g(m,n)\int_\mathcal{M}\delta
 \left(\rho-\rho_{mn}^{(j)}\right)
 \end{eqnarray}
 where the first integral is over the energy interval, and the second is over the domain. The
 function $g(m,n)$ is the quotient of the appropriate Jacobian and
 $\nu_{mn}$. Since in the vicinity of  $\rho=\pi/\sqrt8$ two
 solutions for (\ref{eq:rhorec}) coalesce, we expect the square root
 singularity at $P(\rho\rightarrow \pi/\sqrt8^+)$ to be a universal feature.
\item In addition, since  $\rho(E_m/E_{mn})$ is convex (see equation \ref{eq:rhorec2} and
 figure \ref{fig:rho_of_ei}), the distribution function should be
 monotonically decreasing.
\end{itemize}
This supports the assumption that the properties (i)-(iv) for
$\rho_{rec}$ and $P_{rec}(\rho)$, which were derived in section
\ref{sec:Separable} for the rectangle, are universal features of
$P(\rho)$ for all two dimensional separable surfaces. Moreover, this
model suggest - at least for separable domains - that a geometric
feature of the nodal pattern, i.e. the area-to-perimeter ratio of a
given domain, can be deduced directly from the underlying classical
dynamics.\\

The suggested model is supported by an explicit derivation of the
limiting area-to-perimeter distribution for several separable
domains. In these calculations we approximate eigenfunctions and
eigenvalues by the WKB method. In addition, we approximate sums over
quantum numbers (see eq. \ref{eq:gendis1}) by integrals and neglect
terms of order $E^{-\frac12}$. The error resulting from these
approximations is of the order of $E^{-\frac12}$ and therefore
converge to zero in the
limit.\\
The theme of the derivations is similar. The Hamiltonian
$\mathcal{H}$ for these systems is homogeneous i.e.:
\begin{eqnarray}\label{eq:homog}
&&\mathcal{H}(\lambda m,\lambda n)=\lambda^2\mathcal{H}(m,n)
\end{eqnarray}
This implies that the energy of the state $\psi_{mn}$ can be
expressed as:
\begin{eqnarray}\label{eq:m_f(z)}
\mathcal{H}(m,n)=m^2\mathcal{H}\left(1,\frac nm\right)=m^2\cdot
h\left(\frac nm\right)
\end{eqnarray}
Therefore, integration over the quantum number $m$ becomes trivial.
We will also use the first term in the Weyl series:
$\sharp\{j:E_j<E\}=4\pi E/\mathcal{A}+O(\sqrt E)$ in order to
estimate $N_I$.

\subsection{The disc billiard}\label{subsec:disc}
Equation (\ref{eq:helmholtz}) can be written in polar coordinates as
\begin{eqnarray}\label{eq:phidisc}
&&\left(\frac{\partial^2}{\partial
r^2}+\frac1r\frac{\partial}{\partial r} +\frac
1{r^2}\frac{\partial^2}{\partial\theta^2}+E\right)\psi(r,\theta)=0
\end{eqnarray}
For the disc billiard the boundary condition are: $\psi|_{r=1}=0$.
The eigenfunction and eigenvalues of (\ref{eq:phidisc}) are:
\begin{eqnarray}\label{eq:psidisc}
&&\psi_{mn}(r,\theta)=\cos(m\theta+\varphi)J_m(j_{mn}r)\quad,\quad
E_{mn}={j_{mn}}^2
\end{eqnarray}
where $\varphi$ is an arbitrary phase and $j_{mn}$ is the $n$th zero
of $J_m(r)$. The nodal domains of $\psi_{mn}$ will be $2m$ replicas
(or one for $m=0$) of a slice containing $n$ domains ; we will
enumerate them as $\{\omega_{mn}^{(i)}\}_{i=1\dots n}$, where
$\omega_{mn}^{(1)}$ is the most inner domain. The area and perimeter
of $\omega_{mn}^{(i)}$ are:
\begin{eqnarray}
\mathcal{A}_{mn}^{(i)}=\frac{\pi}{2m}\frac{j_{mi}^2-j_{m,i-1}^2}{j_{mn}^2}\\\nonumber
L_{mn}^{(i)}=\frac{\pi}{m}\frac{j_{mi}+j_{m,i-1}}{j_{mn}}+2\frac{j_{mi}-j_{m,i-1}}{j_{mn}}
\end{eqnarray}
An implicit semi-classical expression for $j_{mn}$ can be deduced by
applying the WKB approximation to (\ref{eq:phidisc}):
\begin{eqnarray}\label{eq:BSdisc}
&&n=\int_\frac m {j_{_{mn}}}^1 \sqrt{j_{mn}^2-{{m^2}\over{r^2}}}dr
\Rightarrow\\ \nonumber && \quad \quad \quad \quad \quad j_{mn}=\pi
A_{mn}\left(n+\left({1\over 2}-C_{mn}\right)m\right)
\end{eqnarray} where:
\begin{eqnarray}\label{eq:discfactors}
&& C_{mn}={1 \over\pi}
\arctan\left(\sqrt{{m^2}\over{j_{mn}^2-m^2}}\right)\\\nonumber
&&A_{mn}=\sqrt{1+{{m^2}\over{\pi^2\left(n+\left({1\over
2}-C_{mn}\right)m\right)^2}}}
\end{eqnarray}
Setting $z={n\over m},z'={i\over m}$, and substituting
(\ref{eq:BSdisc}) into (\ref{eq:discfactors}) we get:
\begin{eqnarray}\label{eq:z(c)}
&&z={{\cot(\pi C_{mn})}\over\pi}+C_{mn}-{1\over 2}
\end{eqnarray}
which implies that $C_{mn}$ depends on $z$ solely. Since $A_{mn}$
varies with $n$ as $E^{-\frac12}$, we can approximate:
$A_{mn}\approx A_{m,n-1}$. Expressing $\rho_{mn}^{(i)}$ in terms of
(\ref{eq:BSdisc},\ref{eq:discfactors}) yields:
\begin{eqnarray}\label{eq:rho(c)}
&&\rho_{mn}^{(i)}=
\frac{\frac{\pi\left({j_{mi}}^2-{j_{m,i-1}}^2\right)}{2m}}
{\frac{\pi}{m}\left(j_{mi}+j_{m,i-1}\right)+2\left(j_{mi}-j_{m,i-1}\right)}\\
\nonumber &&
\quad\quad\quad\quad\quad\quad\quad\quad\quad\quad={\pi\over
2}\sqrt{1\over{1+\sin\left(2\pi C(z')\right)}}
\end{eqnarray}
Bearing in mind that for a point $\vec{r}\in\omega_{mn}^{(i)}$:
\begin{eqnarray}\label{eq:C_of_Em}
E_m(\vec{r})=\frac1{r^2 \psi}\frac{\partial^2\psi}{\partial\theta^2}
=\frac{m^2\cdot j_{mn}^2}{j_{mi}^2}(1+O(n^{-1}))
\end{eqnarray}
It can be shown  that equation (\ref{eq:rho(c)}) is equivalent to
(\ref{eq:rhorec2}).\\
Integrating (\ref{eq:gendis1}) over the variables $m,C \equiv
C(z),C'\equiv C'(z')$, we get:
\begin{eqnarray}\label{eq:Pdisc1}
&&P_I(\rho)=-{8\over{\epsilon g}}\int_I dmdC {1\over z}
{{dz}\over{dC}} \int_C^{1\over 2} \delta\left(\rho-{\pi\over 2}
\sqrt{1\over{1+2\sin(2\pi C')}}\right)dC'
\end{eqnarray}
Performing the integration we get the limiting distribution:
\begin{eqnarray}\label{eq:Pdisc2}
P(\rho)=&&{4\over{\rho\sqrt{8\rho^2-\pi^2}}}\times\\\nonumber
&&\frac{\pi}{2}\left({{4\rho^2+\pi\sqrt{8\rho^2-\pi^2}}\over{4\rho^2-\pi\sqrt{8\rho^2-\pi^2}}}\right.
\int_0^{\gamma_1(\rho)} {{\sin(2\pi C)}\over{1+(C-{1\over
2})\pi\tan(\pi C)}} dC\\\nonumber
&&+\left.{{4\rho^2-\pi\sqrt{8\rho^2-\pi^2}}\over{4\rho^2+\pi\sqrt{8\rho^2-\pi^2}}}
\int_0^{\gamma_2(\rho)} {{\sin(2\pi C)}\over{1+(C-{1\over
2})\pi\tan(\pi C)}}dC\right)
\end{eqnarray}
where:
\begin{eqnarray}
\gamma_1(\rho)={1\over{2\pi}}\arcsin\left({{\pi^2-4\rho^2}
\over{4\rho^2}}\right) \quad ;\quad \gamma_2(\rho)={1\over
2}-\gamma_1(\rho)
\end{eqnarray}

\begin{figure}[h]
\centering
  \includegraphics[width=4in]{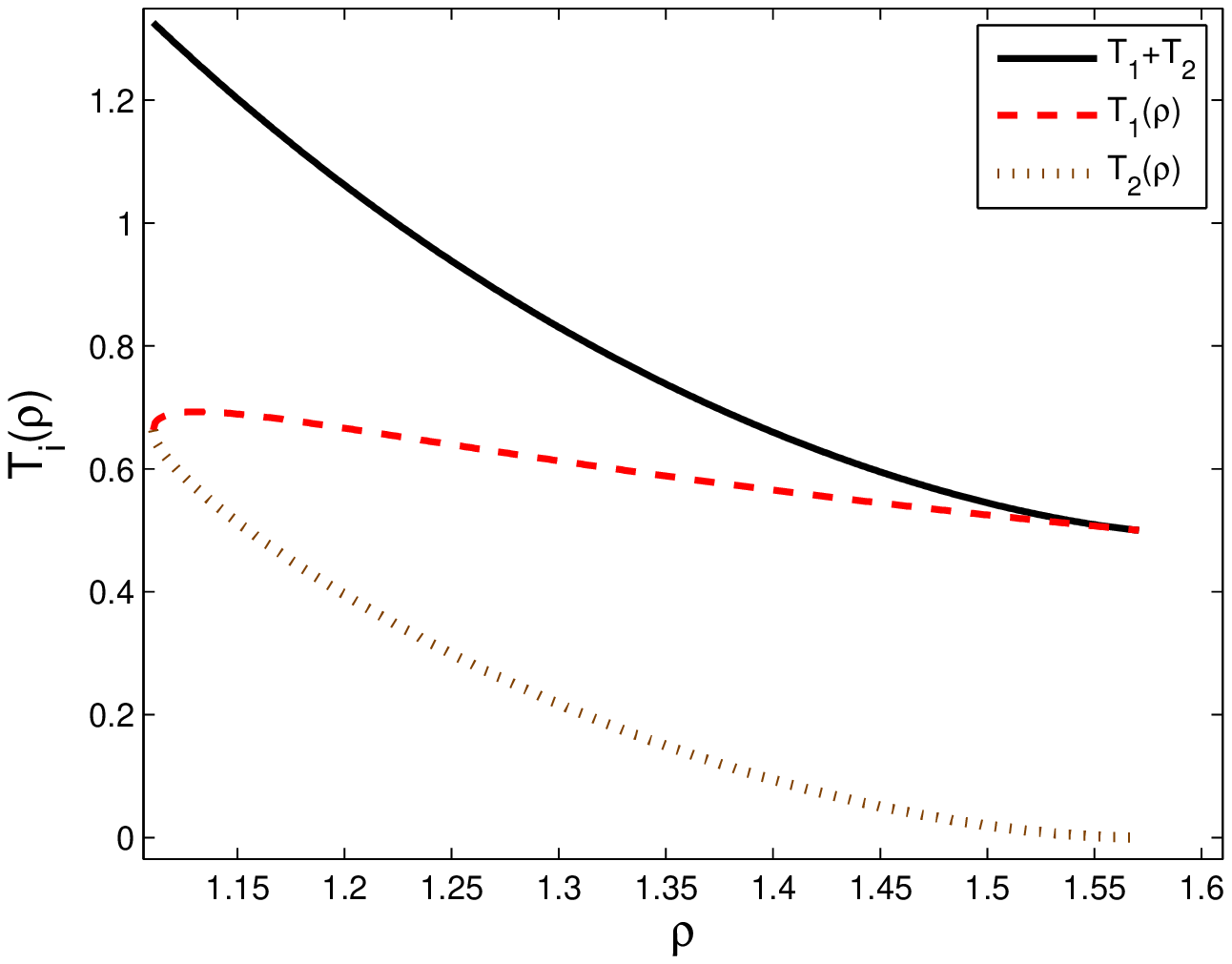}
\caption{A realization of the functions: $T_i(\rho)=c_i(\rho)
\int_{x_i(\rho)}{f(x)dx}$\quad for the disc. $T=T_1+T_2$ is
monotonic, finite and positive.}\label{fig:c1c2}
\end{figure}

\subsection{Surfaces of revolution}\label{subsec:SSR}
The investigated surface of revolution $\mathcal{M}$ is generated by
a rotation of the analytic profile curve $f(x)$ (where $x\in[-1,1]$)
around the $x$ axis. We restrict $f(x)$ by:
$f(x)_{x\rightarrow\pm1}=a_\pm\sqrt{1\mp x}$, which ensures
smoothness at the poles of the surface. In addition, we request that
$f''(x)<0$, so $f(x)$ has a single maximum: $f_{max}\equiv
f(x_{max})$.\\
The Lagrangian of the surface is given by:
\begin{eqnarray}\label{eq:lagrangian}
\mathcal{L}={1\over 2} \left|{ds\over dt}\right|^2= \frac 12
\left({(1+f'(x)^2)\dot{x}^2+f(x)^2 \dot\theta^2}\right)
\end{eqnarray}
From which the action variables can be deduced:
\begin{eqnarray}\label{eq:actionvar}
&&m={1\over{2\pi}}\oint p_\theta d\theta=f(x)^2\dot{\theta}\\
\nonumber &&n={1\over{2\pi}}\oint p_xdx=
{1\over\pi}\int_{x_-}^{x_+}{1\over{f(x)}}\sqrt{(Ef(x)^2-m^2)(1+f'(x)^2)}dx
\end{eqnarray}
where $x_-$ , $x_+$ are the classical turning points which satisfy: $Ef(x_{\pm})^2-m^2=0$.\\
As in \ref{subsec:disc} the nodal domains will be $2m+\delta_{m0}$
copies of a slice with $n$ domains and will be denoted by
$\{\omega_{mn}^{(i)}\}_{i=1}^{n}$. The homogeneity of the
Hamiltonian follows directly from (\ref{eq:actionvar}). We will
follow the notations $z={n \over m},z'={i \over m}$ to get:
\begin{eqnarray}\label{eq:homog1}
&&\Rightarrow E(m,z)=m^2 \mathcal{E}^2(z)
\end{eqnarray}
The WKB approximation to the eigenfunctions is
\begin{eqnarray}\label{eq:PsiSor}
\psi_{mz}(x,\theta)={{\cos(m\theta+\varphi)}\over\sqrt{k}}\times\\\nonumber
&&\left(\cos\left(\int_{x_-}^x kdx-{\pi\over
4}\right)+\cos\left(\int_x^{x_+}kdx-{\pi\over 4}\right)\right)
\end{eqnarray}
where
\begin{eqnarray}\nonumber
&&k={m\over{f(x)}}\sqrt{(\mathcal{E}_z^2 f(x)^2-1)(1+f'(x)^2)}
\end{eqnarray}
In the limit of large $n$, the nodal points density on the curve is
high. Therefore, applying the WKB approximation, successive nodal
points $x_{i-1}$ , $x_{i}$ should satisfy:
\begin{eqnarray}\label{eq:sorquantization}
&&\pi=m\int_{x_{i-1}}^{x_{i}}\sqrt{\mathcal{E}_z^2
f(x)^2-1}{\sqrt{1+f'(x)^2} \over{f(x)}} \\\nonumber
&&\quad\thickapprox
(x_{i}-x_{i-1})m\sqrt{\mathcal{E}_z^2f(x_{i})^2-1}{{\sqrt{1+f'(x_{i})^2}}\over{f(x_{i})}}
\end{eqnarray}
In addition, due to the homogeneity of $\mathcal{H}$, $x_{\pm}$
depend on $z$ solely, therefore:
\begin{eqnarray}\label{eq:xofz}
&&\pi i=m\int_{x_-}^{x_{i}}\sqrt{\mathcal{E}_z^2 f(x)^2-1}
{\sqrt{1+f'(x)^2} \over{f(x)}}dx\Rightarrow \\\nonumber && z'=\frac
1 \pi \int_{x_-}^{x_{i}}\sqrt{\mathcal{E}_z^2
f(x)^2-1}{\sqrt{1+f'(x)^2} \over{f(x)}}dx=z'(x_i,z)
\end{eqnarray}
Therefore, the location of the $i^{th}$ zero will be an (implicit)
function $x_{i}(z,z')$ and will not be depended on $m$. The area and
perimeter of the nodal domains are:
\begin{eqnarray}\label{eq:area_peri}
\mathcal{A}_{mn}^{(i)}=\frac{\pi}{2m}\int_{x_{i-1}}^{x_{i}}f(x)\sqrt{1+f'(x)^2}dx\\\nonumber
L_{mn}^{(i)}=\frac{\pi}{2m}(f(x_{i-1})+f(x_{i}))+\int_{x_{i-1}}^{x_{i}}\sqrt{1+f'(x)^2}dx
\end{eqnarray}
Therefore
\begin{eqnarray}\label{eq:rhosor}
&&\rho_{m,n}^{(i)}={{\pi\int_{x_{i-1}}^{x_{i}}f(x)\sqrt{1+f'(x)^2}dx
\sqrt{E_{nm}}}\over{\pi (f(x_{i-1})+f(x_{i}))+2m
\int_{x_{i-1}}^{x_{i}} \sqrt{1+f'(x)^2}dx}}=\rho(z,z')
\end{eqnarray}
Substituting (\ref{eq:sorquantization}) in (\ref{eq:rhosor}) we get:
\begin{eqnarray}\label{eq:rhosor2}
\rho(z,z')={\pi\over 2}
{{\mathcal{E}(z)f(x_{i})}\over{1+\sqrt{(\mathcal{E}(z)f(x_{i}))^2-1}}}
\end{eqnarray}
where $\mathcal{E}(z)f(x_{i})=1$ at the turning points. Since for
$\vec{r}\in\omega_{mn}^{(i)}$:
\begin{eqnarray}\label{eq:R_of_Em}
E_m(\vec{r})=\frac{m^2}{f^2(x)}=\frac{m^2}{f^2(x_{i})}(1+O(n^{-1}))
\end{eqnarray}
Equation (\ref{eq:rhosor}) is equivalent to (\ref{eq:rhorec2}) as well.\\
Integrating (\ref{eq:gendis1}) over $m,z,z'$ we get:
\begin{eqnarray}\label{eq:dister}
&&P_I(\rho)= {{4\pi}\over{g\epsilon |\mathcal{M}|}} \int_0^\infty dz
\int_{\frac{\sqrt{\epsilon}}{\mathcal{E}(z)}}^{\frac{\sqrt{\epsilon}(1+g)}
{\mathcal{E}(z)}}dm {m\over
z}\int_0^z\delta(\rho-\rho(z'z))dz'\\
\nonumber &&\quad\quad\quad ={{2\pi}\over{|\mathcal{M}|}}
\int_0^{\infty} {{dz}\over{z \mathcal{E}(z)}}
\int_o^z\delta(\rho-\rho(z'z))dz'
\end{eqnarray}
Setting $\alpha_{z'z}=\mathcal{E}(z)f(x_{i}')$, we get that for
$\rho\in [\pi/\sqrt8,\pi/2]$ and fixed $z$, there are two allowed
values of $\alpha$:
\begin{eqnarray}\label{eq:alpha}
\alpha_{1,2}={{2\rho}\over{\pi^2-4\rho^2}}\left(\pi\pm\sqrt{8\rho^2-\pi^2}\right)
\end{eqnarray}
Since $f^{-1}(x)$ is doubly valued, there are four allowed values of
$z'$:
\begin{eqnarray}\label{eq:z'}
z'_{j1}=\min\left(f^{-1}\left({{\alpha_j}\over{\mathcal{E}(z)}}\right)\right)\quad
,\quad
z'_{j2}=\max\left(f^{-1}\left({{\alpha_j}\over{\mathcal{E}(z)}}\right)\right)
\end{eqnarray}
for $j=1,2$. Therefore
\begin{eqnarray}\label{eq:Psor3}
P(\rho)={{2\pi}\over{|\mathcal{M}|}} \int_0^\infty{{dz}\over{z
\mathcal{E}(z)^2}}\sum_i\int_0^z{{\delta(z'-z'_i)}\over{\rho_{z'z}'(z')}}dz'=
\\\nonumber P_{rec}(\rho)\cdot \left(T_1(\rho)+T_2(\rho)\right )
\end{eqnarray}
where
\begin{eqnarray}\nonumber
T_1=\frac{2\pi^2\rho{{(4\rho^2-\pi\sqrt{8\rho^2-\pi^2})^2}}}
{|\mathcal{M}|(\pi^2-4\rho^2)^2(\pi-\sqrt{8\rho^2-\pi^2})}
\int_{z'_{1i}}^\infty\sum_{i=1,2}{{dz}\over{z
\mathcal{E}(z)^3}}\left|{{df(x_{z'z}')}\over{dz'}}\right|^{-1}\\\nonumber
T_2=\frac{2\pi^2\rho{{(4\rho^2+\pi\sqrt{8\rho^2-\pi^2})^2}}}
{|\mathcal{M}|(\pi^2-4\rho^2)^2(\pi+\sqrt{8\rho^2-\pi^2})}
\int_{z'_{2i}}^\infty\sum_{i=1,2}{{dz}\over{z
\mathcal{E}(z)^3}}\left|{{df(x_{z'z}')}\over{dz'}}\right|^{-1}
\end{eqnarray}
$\mathcal{E}_z$ is a monotonic increasing function, therefore the
integrals in (\ref{eq:Psor3}) diverges at $z=0$. For $z>>1$ ,
$\mathcal{E}(z)=O(z)$, therefore the
integrals converge at infinity.\\
For $\rho\rightarrow\pi/\sqrt8^+$, the coefficient of the integrals
in (\ref{eq:Psor3}) has a square root singularity, while the
integral is converging to a finite positive value, therefore
$P(\pi/\sqrt8+\delta)\sim 1/\sqrt{\delta}$.\\
For $\rho\rightarrow\pi/2-\delta$, the first coefficient in
(\ref{eq:Psor3}) is of the order of $\delta$. The lower limit of
integration is defined by:
\begin{eqnarray}
f_z(z')={{1+\delta}\over{\mathcal{E}_z}}=f_z(0)+{\delta\over{\mathcal{E}_z}}
\Rightarrow\frac{df(z')}{dz'}={\delta\over{z'\mathcal{E}_z}}
\end{eqnarray}
Consequently, the first term in (\ref{eq:Psor3}) will be of order:
\begin{eqnarray}\label{eq:Iterm}
&&I\sim \delta\int_{z'}^\infty {{z'F(z')dz}\over{\delta z
\mathcal{E}_z^3}}
\end{eqnarray}
The value of $z'$ depends on the profile curve $f(x)$, however for
$z'\rightarrow 0$, $I\sim z'\log(z')$, while for
$z'\rightarrow\infty$, we get that $I\sim 1/z'^3$,
therefore $I$ is bounded for all possible values of $z'$.\\
The value of the second term is due to contributions of nodal
domains which satisfy
\begin{eqnarray}\label{eq:IIterm}
&&\rho_{z'z}={\pi\over 2}-\delta \Rightarrow
\mathcal{E}_z\thicksim{1\over{f_z(z')\delta}}\Rightarrow z=\frac
nm\geqq O(\frac1\delta)
\end{eqnarray}
Therefore, only eigenfunctions for which $n/m\ge O(1/\delta)$
contribute to the second term, as a result, it is bounded by:
\begin{eqnarray}\label{eq:IIterm2}
{{\int_I dndm \theta(n-m\delta^{-1})}\over{\int_I dndm}} \cong
\frac{\delta g\epsilon}{\frac{|\mathcal{M}|}{4\pi}g\epsilon}
\thicksim\delta
\end{eqnarray}\\
therefore $P(\pi/2-\delta)$ is finite, and the universal features
specified in section \ref{sec:Separable} are all fulfilled by
(\ref{eq:Psor3}).\\

\section{Numerical methods for evaluation of the perimeter length}
In order to evaluate the area-to-perimeter ratios and their
distribution for the random-wave ensemble and chaotic billiards, we
have simulated the appropriate wavefunctions on a grid.\\
We have calculated the statistics for 5000 realizations of random
waves, where in each realization we summed over 70 terms in
(\ref{eq:randomwave}); For chaotic billiards we have reproduced the
first 2430 eigenfunctions of a Sinai Billiard and
the first 2725 eigenfunctions of a stadium billiard.\\
The (seemingly simple) task of perimeter evaluation must be carried
out carefully; it can be shown that naive methods, like perimeter's
pixels counting, produce an error which is independent of the
sampling resolution. In order to avoid this error, we have
approximated the nodal line as a polygon, where the vortices are
calculated using a linear approximation. We have set the sampling
resolution to contain 85 pixels along the average distance between
two nodal lines ($\sqrt{2}\pi/k$). This resolution was proved to
produce an error which is less then a percent. The measured
perimeter is expected to be shorter then the real one, as we are
approximating a curve by a polygon.\\
The accuracy of the method was tested by calculating the ratio
between $L_T$-the total length of the nodal set, and the area. It is
known that for the random-wave ensemble \cite{Berry02}:
\begin{eqnarray}\label{eq:total length}
\frac{\langle L_T\rangle}{A\cdot k}=\frac1{2\sqrt{2}}
\end{eqnarray}
The numerical values for this ratio were between:
\begin{eqnarray}\label{eq:numerics}
\frac{0.998}{2\sqrt{2}}<{\frac{L_T}{A\cdot
k}}^{(M)}<\frac{1.003}{2\sqrt{2}}
\end{eqnarray}
It should be noted that when we calculate the total nodal length, we
have to calculate the perimeter of nodal domains at the edge of the
grid. In many cases (i.e. where the edge domains are very small) the
perimeter calculated for them is larger then the real value. It
seems likely that the error due to this effect is of the order of
the error due to polygonal approximation, therefore the two
compensate each other, to yield a total error which is relatively
small, of order $L_T\cdot 10^{-3}$.\\

\noindent {\bf{Bibliography}}

\begin{thebibliography}{10}

\bibitem{Blum}
G.~{Blum}, S.~{Gnutzmann}, and U.~{Smilansky}.
\newblock Nodal domains statistics: A criterion for quantum chaos.
\newblock {\em Physical Review Letters}, 88(11):114101, March 2002.

\bibitem{Berry77}
M.~V. Berry.
\newblock Regular and irregular semiclassical wave functions.
\newblock {\em Journal of Physics A Mathematical General}, 10:2083--2091, 1977.

\bibitem{Handy}
R.~M. Stratt, N.~C. Handy, and W.~H. Miller.
\newblock On the quantum mechanical implications of clasical ergodicity.
\newblock {\em Jour. of Chem. Phys.}, 71:3311--3322, October 1979.

\bibitem{Donnelly}
H.~Donnelly and C.~Fefferman.
\newblock Nodal sets of eigenfunctions on reimannian manifolds.
\newblock {\em Inventiones Mathematicae}, pages 161--183, February 1988.
\newblock 10.1007/BF01393691.

\bibitem{Bruning}
J.~Br{\"u}ning.
\newblock \"uber knoten von eigenfunktionen des laplace-beltrami-operators.
\newblock {\em Math. Z.}, 158(1):15--21, 1978.

\bibitem{Berry02}
M.~V. Berry.
\newblock Statistics of nodal lines and points in chaotic quantum billiards:
  perimeter corrections, fluctuations, curvature.
\newblock {\em Journal of Physics A Mathematical General}, 35:3025--3038, April
  2002.

\bibitem{Monastra}
A.~G. {Monastra}, U.~{Smilansky}, and S.~{Gnutzmann}.
\newblock {Avoided intersections of nodal lines }.
\newblock {\em Journal of Physics A Mathematical General}, 36:1845--1853,
  February 2003.

\bibitem{Bogomolny02}
E.~{Bogomolny} and C.~{Schmit}.
\newblock {Percolation Model for Nodal Domains of Chaotic Wave Functions}.
\newblock {\em Physical Review Letters}, 88(11):114102, March 2002.

\bibitem{Bogomolny06}
E.~Bogomolny, R.~Dubertrand, and C.~Schmit.
\newblock {S}{L}{E} description of the nodal lines of random wave functions,
  2006.

\bibitem{Smirnov}
S.~Smirnov.
\newblock Critical percolation in the plane: conformal invariance, cardy's
  formula, scaling limits.
\newblock {\em C. R. Acad. Sci. Paris S\'er. I Math.}, 333(3):239--244, 2001.

\bibitem{Foltin04}
G.~Foltin, S.~Gnutzmann, and U.~Smilansky.
\newblock The morphology of nodal lines random waves versus percolation.
\newblock {\em Journal of Physics A Mathematical General}, 37:11363--11371,
  November 2004.

\bibitem{Freitas}
P.~Freitas and P.~Antunes.
\newblock New bounds for the principal dirichlet eigenvalue of planar regions.
\newblock {\em Exp. Math.}, 15:333--342, 2006.

\bibitem{Courant}
R.~Courant and D.~Hilbert.
\newblock {\em Methods of mathematical physics. {V}ol. {I}}.
\newblock Interscience Publishers, Inc., New York, N.Y., 1953.

\bibitem{Makai}
E.~Makai.
\newblock On the principal frequency of a membrane and the torsional rigidity
  of a beam.
\newblock In {\em Studies in mathematical analysis and related topics}, pages
  227--231. Stanford Univ. Press, Stanford, Calif., 1962.

\bibitem{Polya}
G.~P{\'o}lya.
\newblock Two more inequalities between physical and geometrical quantities.
\newblock {\em J. Indian Math. Soc. (N.S.)}, 24:413--419 (1961), 1960.

\bibitem{Bleher}
P.~Bleher.
\newblock Distribution of energy levels of a quantum free particle on a surface
  of revolution.
\newblock 1993.

\bibitem{Stauffer}
D.~Stauffer and A.~Aharony.
\newblock {\em Introduction to percolation theory, 2nd edition}.
\newblock Taylor \& Francis Ltd., London, 1994.

\bibitem{Mazzolo}
A.~Mazzolo, B.~Roesslinger, and W.~Gille.
\newblock Properties of chord length distributions of nonconvex bodies.
\newblock {\em Journal of Mathematical Physics}, 44:6195--6208, December 2003.

\bibitem{Rice}
S.~O. {Rice}.
\newblock {Mathematical Analysis of Random Noise}.
\newblock {\em Bell Systems Tech.~J., Volume 23, p.~282-332}, 23:282--332,
  1944.

\end{thebibliography}


\begin{thebibliography}{10}



\bibitem{Stockmann} H.-J.~St\"ockmann,
  {\em Introduction to Quantum Chaos},
  (Cambridge University Press, Cambridge, UK, 1999).

\bibitem{Haake} F.~Haake,
  {\em Quantum Signatures of Chaos} (2nd~ed., Springer, Berlin, 2000).

\bibitem{Blum}
  G.~{Blum}, S.~{Gnutzmann}, and U.~{Smilansky},
  Phys.~Rev.~Lett. \textbf{88}, 114101 (2002).

\bibitem{Guhr} T.~Guhr, 
A.~M\"uller-Groeling and H.A.~Weidenm\"uller,
Phys.~Rep. \textbf{299}, (189).

\bibitem{Berry77}
  M.~V. Berry.
  J.~Phys.~A \textbf{10}, 2083 (1977).

\bibitem{Handy}
  R.~M. Stratt, N.~C. Handy, and W.~H. Miller,
  J.~Chem.~Phys. \textbf{71}, 3311
  (1979).

\bibitem{Donnelly}
  H.~Donnelly and C.~Fefferman,
  Invent.~Math. \textbf{93}, 161 (1988).

\bibitem{Bruning}
  J.~Br{\"u}ning, Math.~Z. \textbf{158}, 15 (1978).

\bibitem{Berry02}
  M.~V. Berry,
  J.~Phys.~A \textbf{35}, 3025 (2002).

\bibitem{Monastra}
  A.~G. {Monastra}, U.~{Smilansky}, and S.~{Gnutzmann}.
  J.~Phys.~A \textbf{36}, 1845
  (2003).

\bibitem{Bogomolny02}
  E.~{Bogomolny} and C.~{Schmit},
  Phys.~Rev.~Lett. \textbf{88}, 114102 (2002)

\bibitem{keating}
  J.~Keating, J.~Marklof, and I.G.~Williams,
  Phys.~Rev.~Lett. \textbf{97},034101 (2006).

\bibitem{williams} I.G.~Williams, PhD thesis (Bristol, 2006).

\bibitem{Bogomolny06}
  E.~Bogomolny, R.~Dubertrand, and C.~Schmit,
  J.~Phys.~A \textbf{40}, 381
  (2007).

\bibitem{Smirnov}
  S.~Smirnov,
  C.~R.~Acad.~Sci.~Paris~S\'er.~I~Math. \textbf{333}, 239 (2001).

\bibitem{Foltin04}
  G.~Foltin, S.~Gnutzmann, and U.~Smilansky,
  J.~Phys. A  
  \textbf{37}
  11363, 2004.

\bibitem{Freitas}
  P.~Freitas and P.~Antunes,
  Exp.~Math. \textbf{15}, 333 (2006).

\bibitem{Courant}
  R.~Courant and D.~Hilbert,
  {\em Methods of mathematical physics. {V}ol. {I}}.
  (Interscience Publishers, New York, 1953).

\bibitem{Makai}
  E.~Makai,
  in G.~Szego (editor) {\em Studies in mathematical analysis and related topics, Essays in honor of George Polya}, pages
  227--231 (Stanford Univ. Press, Stanford, Calif., 1962).

\bibitem{Polya}
  G.~P{\'o}lya,
  J.~Indian~Math.~Soc. (N.S.) \textbf{24}, 413 (1960).

\bibitem{Osserman}
  R.~ Osserman, Comment.~Math.~Helvetici
  \textbf{52}, 545 (1977).

\bibitem{Aiba}
  H.~Aiba and T.~Suzuki,
  Phys.~Rev. E \textbf{72}, 066214 (2005).
  
\bibitem{Bleher}
  P.~Bleher, Duke Math.~J.~\textbf{74}, 45 (1994).

\bibitem{Stauffer}
  D.~Stauffer and A.~Aharony,
  {\em Introduction to percolation theory}
  (2nd ed., Taylor \& Francis Ltd., London, 1994).

\bibitem{Mazzolo}
  A.~Mazzolo, B.~Roesslinger, and W.~Gille,
  J.~Math.~Phys. \textbf{44}, 6195 (2003).

\bibitem{Rice}
  S.~O. {Rice}.
  Bell Systems Tech.~J. \textbf{23}, 282 (1944).

\end{thebibliography}

\end{document}